\documentclass[pdftex,twocolumn,epjc3]{svjour3}   
\RequirePackage[T1]{fontenc}

\smartqed  

\RequirePackage{graphicx}
\RequirePackage{mathptmx}      
\RequirePackage{flushend}
\RequirePackage[numbers,sort&compress]{natbib}
\RequirePackage[colorlinks,citecolor=blue,urlcolor=blue,linkcolor=blue]{hyperref}
\RequirePackage{amsmath}
\RequirePackage{multirow}
\RequirePackage{xspace}
\RequirePackage{xcolor}
\RequirePackage{cancel}
\RequirePackage[caption=false]{subfig}
\RequirePackage{subfig}
\RequirePackage{blindtext, subfig}
\RequirePackage{amsmath}
\usepackage{amssymb}
\usepackage{dblfloatfix}
\usepackage{lipsum}
\usepackage{ulem}
\RequirePackage{fix-cm}

\RequirePackage{comment}
\RequirePackage{hyperref}
\RequirePackage{slashed}
\RequirePackage[caption=false]{subfig}
\RequirePackage{tabu}
\usepackage{hyperref}
\usepackage{multicol}
\usepackage[super]{nth}
\usepackage{booktabs}
\usepackage{dutchcal} 
\begin{document}

\title{Feasibility of the observation of a heavy scalar through the fully hadronic final state at the LHeC}

\author{Elias Malwa\thanksref{e1,addr1}
        \and
        Mukesh Kumar\thanksref{e2,addr1}
        \and Bruce Mellado\thanksref{e4,addr1,addr2}
        \and
        Xifeng Ruan\thanksref{e5,addr1}
}
\thankstext{e1}{e-mail: \textcolor{blue}{elias.malwa@cern.ch}}
\thankstext{e2}{e-mail: \textcolor{blue}{mukesh.kumar@cern.ch}}
\thankstext{e4}{e-mail: \textcolor{blue}{bmellado@mail.cern.ch}}
\thankstext{e5}{e-mail: \textcolor{blue}{xifeng.ruan@cern.ch}}

\institute{School of Physics and Institute for Collider Particle Physics, University of the Witwatersrand, Johannesburg, Wits 2050, South Africa. \label{addr1}
           \and
           iThemba LABS, National Research Foundation, PO Box 722, Somerset West 7129, South Africa. \label{addr2}
}

\date{Received: date / Accepted: date}

\maketitle

\begin{abstract}
The proposed future Large Hadron Electron Collider provides sufficient center of mass energies, $\sqrt{s}$, to probe heavy particles decaying into $W^\pm(Z)-$boson of mass $>2m_W$ $(2m_Z)$. In this work we present a study to produce one such heavy $C P$ even scalar $H$ of mass $2m_h < m_H < 2 m_t$ through charged-current production mode where $H \to W^+W^-$, where hadronic decay of $W^\pm-$boson is considered to reconstruct $m_H$. Due to the presence of missing energy and forward jet in this channel, it is challenging to reconstruct $m_H$ with above final state and thus we employed three different reconstruction methods and discuss the significance of each one. For this analysis we consider a benchmark value of $m_H = 270$\,GeV and $\sqrt{s} \approx 1.3$\,TeV with an assumed luminosity of 1\,ab$^{-1}$.
\end{abstract}

\section{Introduction}
\label{intro}

To date many existing models beyond the Standard Model (BSM) like the two-Higgs doublet models~\cite{Branco:2011iw} and its extensions incorporates scalars of mass lower or higher than the SM Higgs-boson ($m_h = 125$\,GeV)~\cite{ATLAS:2012yve} with models parameters heavily constrained by existing experimental data and theoretical limits. The multi-lepton anomalies seen in {Run~1} data at ATLAS and CMS are explained in a two-Higgs doublet model with additional real singlet scalar (2HDM+S)~\cite{vonBuddenbrock:2015ema,vonBuddenbrock:2017gvy, vonBuddenbrock:2016rmr,Buddenbrock:2019tua,vonBuddenbrock:2020ter,Hernandez:2019geu}.\footnote{For a recent review of anomalies see Ref.~\cite{Fischer:2021sqw}.} In this model the mass of the heaviest $CP$-even scalar $H$ is considered in the interval $2m_h \leq m_H < 2m_t$, where $m_t$ is the mass of top-quark. {The 2HDM+S model with different mass ranges of scalars are also well motivated from theories BSM \cite{Muhlleitner:2016mzt,Muhlleitner:2017dkd,Krause:2017mal,Engeln:2018mbg,Ferreira:2019iqb}, possibilities of existence of BSM scalars at the Large Hadron Collider data~\cite{Arhrib:2018qmw,Biekotter:2021qbc} and future $e^+e^-$ collider~\cite{Azevedo:2018ubr,Azevedo:2018llq}, to explain dark matter abundance~\cite{Glaus:2022rdc,Azevedo:2021ylf,Engeln:2020fld}, di-Higgs production~\cite{Abouabid:2021yvw}, excess seen at 96~GeV~\cite{Biekotter:2019kde,Heinemeyer:2021msz} and to explain recent CDF~\cite{CDF:2022hxs} $W$-mass measurements~\cite{Biekotter:2022abc}. Heavy scalars searches in $WW/ZZ$ channels are considered at CMS and ATLAS~\cite{CMS:2022bcb,ATLAS:2017otj,ATLAS:2020tlo}. The discovery potential of heavy Higgs-boson through the resonant di-Higgs production in HL-LHC and FCC-hh has been studied with $4\tau$ and $bb\gamma\gamma$ channels in ``xSM" model~\cite{FCC:2018vvp}.\footnote{The ``xSM" model is the extension of the SM scalar sector with a single real singlet scalar.} Even the physics of dark matter and axions or axions like particles can be connected with $CP$-even or odd scalars~\cite{Bell:2016ekl,Bell:2017rgi,Ghebretinsaea:2022djg}.}    

\begin{figure}[t]
    \includegraphics[width=0.45\textwidth]{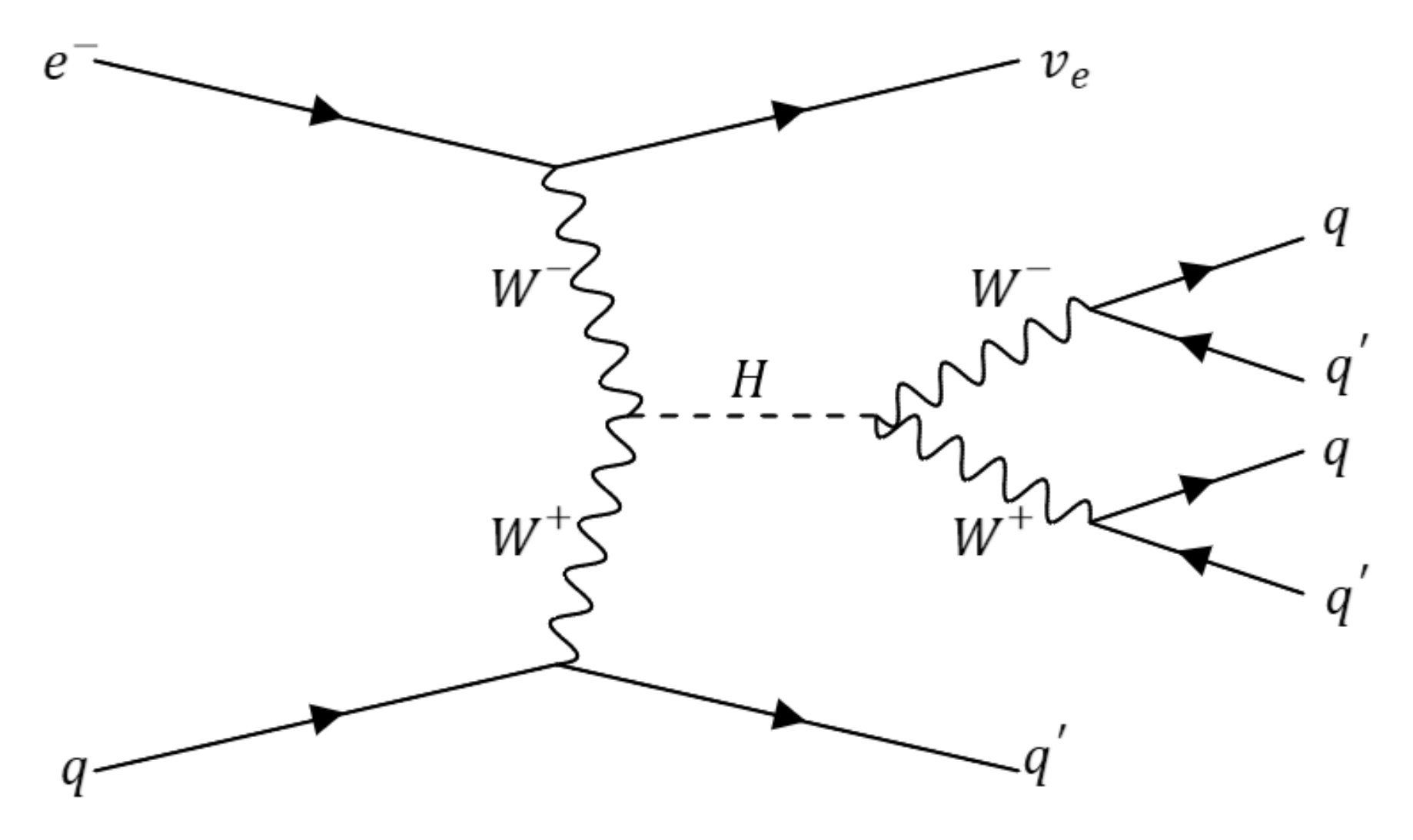}
  \caption{Leading order diagram for signal process $p e^- \to \nu_e H j$, $H \to W^{+}W^{-}$, $W^\pm \to jj$. Here, $q \equiv u, c, \bar{d}, \bar{s}$ and $q' \equiv d, s, \bar{u}, \bar{c}$.}
   \label{fig:Feynman}
\end{figure}

In this work we investigate the possibility of probing $H$ at the {proposed} future electron-proton colliders $via$ the deep-inelastic scattering charged-current (\texttt{CC}) process. The proposed Large Hadron Electron Collider (LHeC) facility at CERN provides sufficient center of mass energy $\sqrt{s} \approx 1.3$\,TeV following electron (proton) energy of $E_{e(p)} = 60$\,GeV (7\,TeV) to explore the allowed mass range of $H$. Interestingly with this mass range one can explore the resonance $H$ via its decay to $W^\pm$ and $Z-$bosons. In this work we consider $H\to W^+W^-$, where $W^\pm$ decay to hadronic final states. However, the mass reconstruction of $H$ through this final state is challenging due to (a) the $W^\pm-$boson emanating from heavy $H$ is boosted with respect to the laboratory system, and hence the jets coming from $W^\pm$ are collimated, and (b) in the $e^- p$ production process, the scattered jet from the proton-line is not easily distinguishable from jets coming from $W^\pm$ (Fig.~\ref{fig:Feynman}). However, the high rapidity ($\eta_j$) region of the scattered jets can be exploited to reconstruct the signal. We also employ a machine learning approach to distinguish the signal and potential backgrounds in this work.

In sect.~\ref{model} we discuss the framework needed to perform this analysis. A description of event simulation and tools needed are discussed in sect.~\ref{Methodology}. The mass reconstruction methods are described in sect.~\ref{massrec}. Summary and discussion of this work is presented in sec.~\ref{Conclusion}.

\section{Model}
\label{model}
To investigate the discovery potential of heavy Higgs boson of mass $2m_h \leq m_H < 2m_t$ in $e^- p$ environment, we consider a model where $H$ corresponds to a real singlet scalar field $\Phi_H$ which mixes with the SM $SU(2)$ doublet Higgs field $\Phi$. Then the Higgs-boson Lagrangian will be modified and can be written as~\cite{Schabinger:2005ei,Dawson:2009yx,Dawson:2017jja}:
\begin{align}
{\cal L}_{\rm Higgs} =&\, (D_\mu\Phi)^2 + (\partial_\mu\Phi_H)^2 + \mu_h^2 \left|\Phi\right|^2 - \lambda_h \left|\Phi\right|^4 \notag \\
&\,  + \mu_H^2 \left|\Phi_H\right|^2 - \lambda_H \left|\Phi_H\right|^4 + \xi  \left|\Phi\right|^2\left|\Phi_H\right|^2. \label{lhiggs} 
\end{align}
In general the, parameters $\mu_h, \mu_H, \lambda_h$ and $\lambda_H$ are all positive in order to have stable potential but $\xi$ may not require any particular sign. We assume that in the above Lagrangian the scalar fields acquire a vacuum expectation values and hence the component fields can be written as:
\begin{align}
\Phi = \frac{1}{\sqrt{2}}
\begin{pmatrix}
G^\pm \\
\phi + {\rm v} + i G^0
\end{pmatrix},
\Phi_H = \frac{1}{\sqrt{2}}\left(\phi_H + {\rm v}_H + i G^\prime \right). \label{vev}
\end{align}
Here the fields $G$ are Goldstone bosons absorbed by the vector bosons, and so no physical pseudoscalar states are left in the spectrum. But the scalar spectrum has two physical states $h$ and $H$ rather than just one of the SM. Also since the singlet do not couple to the $SU(2)_L \times U(1)_Y$ gauge bosons, they do not contribute to $m_W$ and $m_Z$ and hence {\rm v} must take the SM value ${\rm v} = 246$\,GeV. We can also redefine the coefficient of eq.~\ref{lhiggs} such that ${\rm v}_H = 0$. Note that we are not imposing any extra possible symmetries like $\mathbb{Z}_2$ in the scalar sector, and in general $\phi$ will mix with the $\phi_H$ to form the mass eigenstates. We assume the masses of $h$ and $H$ as in previous case, $m_h < m_H$, where $m_h = 125$\,GeV is taken as the SM Higgs boson and $m_H$ as mass of the heavy scalar singlet. The mass eigenstates $h$ and $H$ are related to the gauge eigenstates $\phi$ and $\phi_H$ by a $2\times 2$ unitary matrix\footnote{In general, a $2\times 2$ unitary matrix $V$ can be formed with one parameter $\theta$ as:
\begin{align}
V = \begin{pmatrix}
\cos\theta & \sin\theta \\
-\sin\theta & \cos\theta
\end{pmatrix}\equiv\begin{pmatrix}
V_{11} & V_{12} \\
-V_{12} & V_{11}
\end{pmatrix}, \nonumber
\,\text{where}\, \left|V_{11}\right|^2 + \left|V_{12}\right|^2 = 1.
\end{align}
} 
$V$:
\begin{align}
\begin{pmatrix}
\phi \\
\phi_H
\end{pmatrix}
= V
\begin{pmatrix}
h \\
H
\end{pmatrix}.
\end{align}
Hence the couplings of the gauge bosons and fermions with $h$ will be same as in the SM if $\left|V_{11}\right| = 1$ which implies $\left|V_{12}\right| = \sqrt{1 - \left|V_{11}\right|^2} = 0$. {However in this work we considered $\left|V_{11}\right| \neq 1$ and $\left|V_{12}\right| \neq 0$} . Then the production rates of the $h$ and $H$ are suppressed by a factor $\left|V_{1i} \right|^2$ relative to the SM $h$ production rates. The branching ratios (BRs) of $h$ to the SM particles are identical to the SM BRs, while the BRs of heavy $H$ depend on whether the channel $H \to h h$ are kinematically accessible. For our analysis we scale the $HW^+W^-$ coupling with respect to the SM Higgs boson $hW^+W^-$ coupling. 

\begin{table}[t]
\centering
 \begin{tabular}{lc}
 \toprule
 Process  &  Cross section (fb)  \\
 \toprule
 signal     & 0.49 \\
 \hline
 $e^{-}W^{+}W^{-}j$   &  26.7    \\   
 $e^{-}ZZj$   & 0.13 \\
 $\nu_{e}W^{+}W^{-}j$   &  7.66    \\   
 $\nu_{e}ZZj$   & 2.54  \\ 
 \toprule
\end{tabular}
\caption{Total cross-sections (in fb) for signal production (see text) and potential backgrounds with $E_e = 60$\,GeV and $E_p = 7$\,TeV. The polarisation of $e^-$ is taken to be $-80$\%. The first row represents the signal process and the other four rows are for the dominant background processes.} 
\label{cross_section_table}
\end{table} 
 \begin{table*}[t]
 \centering
 \begin{tabular}{lcccccccc}
 \toprule
   cuts & signal ($S$) & $e^{-}WW+j$ & $e^{-}ZZ+j$ & $v_{e}WW+j$ &$v_{e}{ZZ}+j$& Total Background ($B$) & $S/\sqrt{B}$ & $\sigma(\delta_{sys})$ \\
 \toprule
 initial & 499 & 2680 & 128 & 7660 & 2540 & 13008 & 4.4 & 1.8 \\ \hline 
 at least $5j$  & 211 & 264  & 20  & 1390  & 568  & 2242  & 4.5 & 3.2\\ \hline
 $E^{miss}_{T} > 20$\,GeV  & 182  & 52 & 4 & 1330 & 542 & 1928  & 4.1 & 3.1\\ \toprule
\end{tabular}
  \caption{A summary table of event selections. In the first column the selection criteria are given. The second column contains the weight of the signal process $p e^- \to \nu_e H j$, $H \to W^+W^-$, $W^\pm \to jj$ for $m_H = 270$\,GeV. From column third to sixth dominant weights for backgrounds are given. Seventh column is weighted total number of backgrounds. All weights are calculated with ${\cal L}$ = 1\,ab$^{-1}$. The significance of signal over total background is given in { the eight} column. {In the last column significance with $\delta_{sys} = 2\%$ is estimated.} }
\label{opt}
\end{table*}
\begin{figure*}[t]
   \centering
    \subfloat[\label{njet}]{\includegraphics[width=0.45\textwidth]{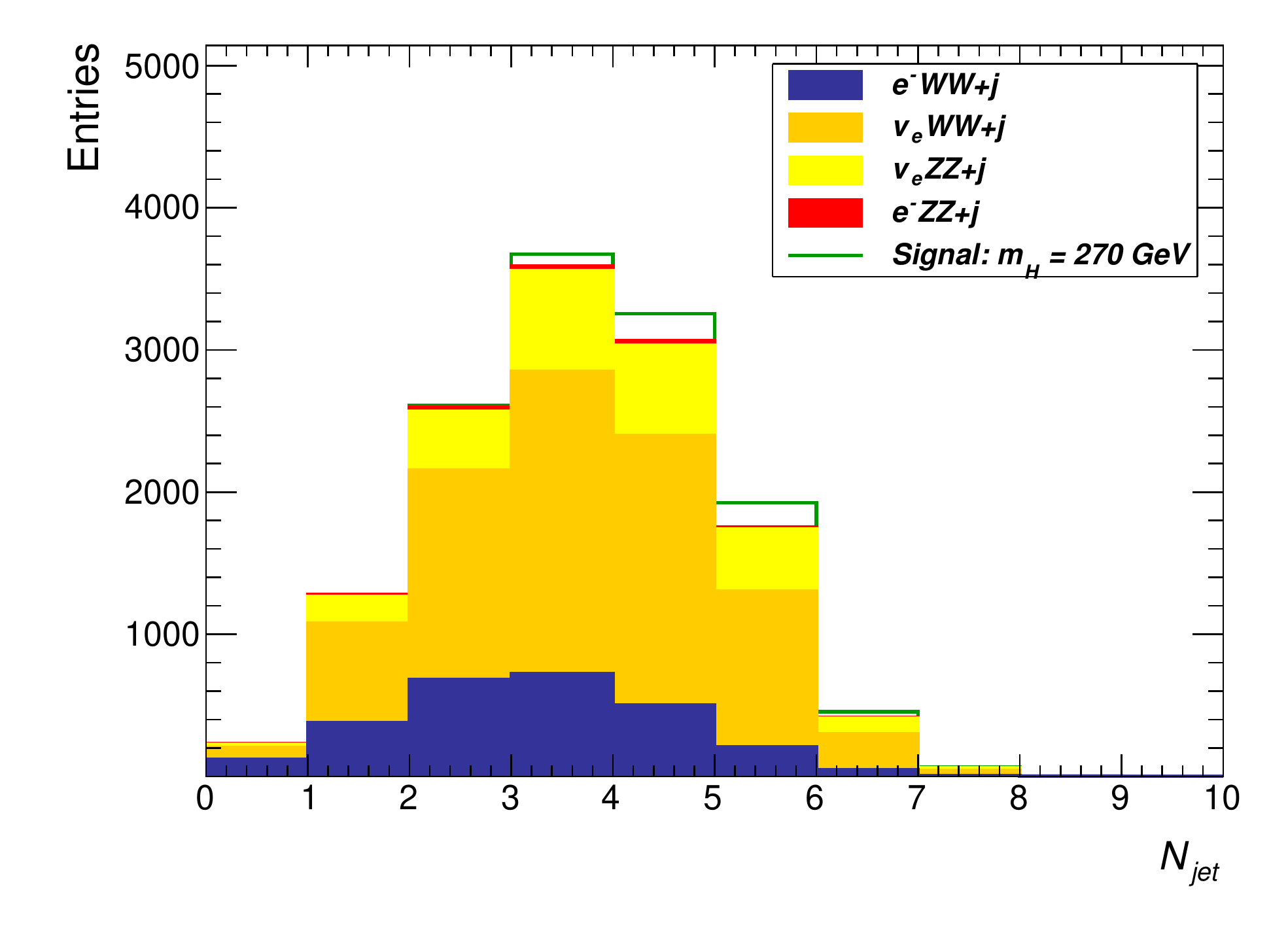}} 
    \subfloat[\label{fwjet}]{\includegraphics[width=0.45\textwidth]{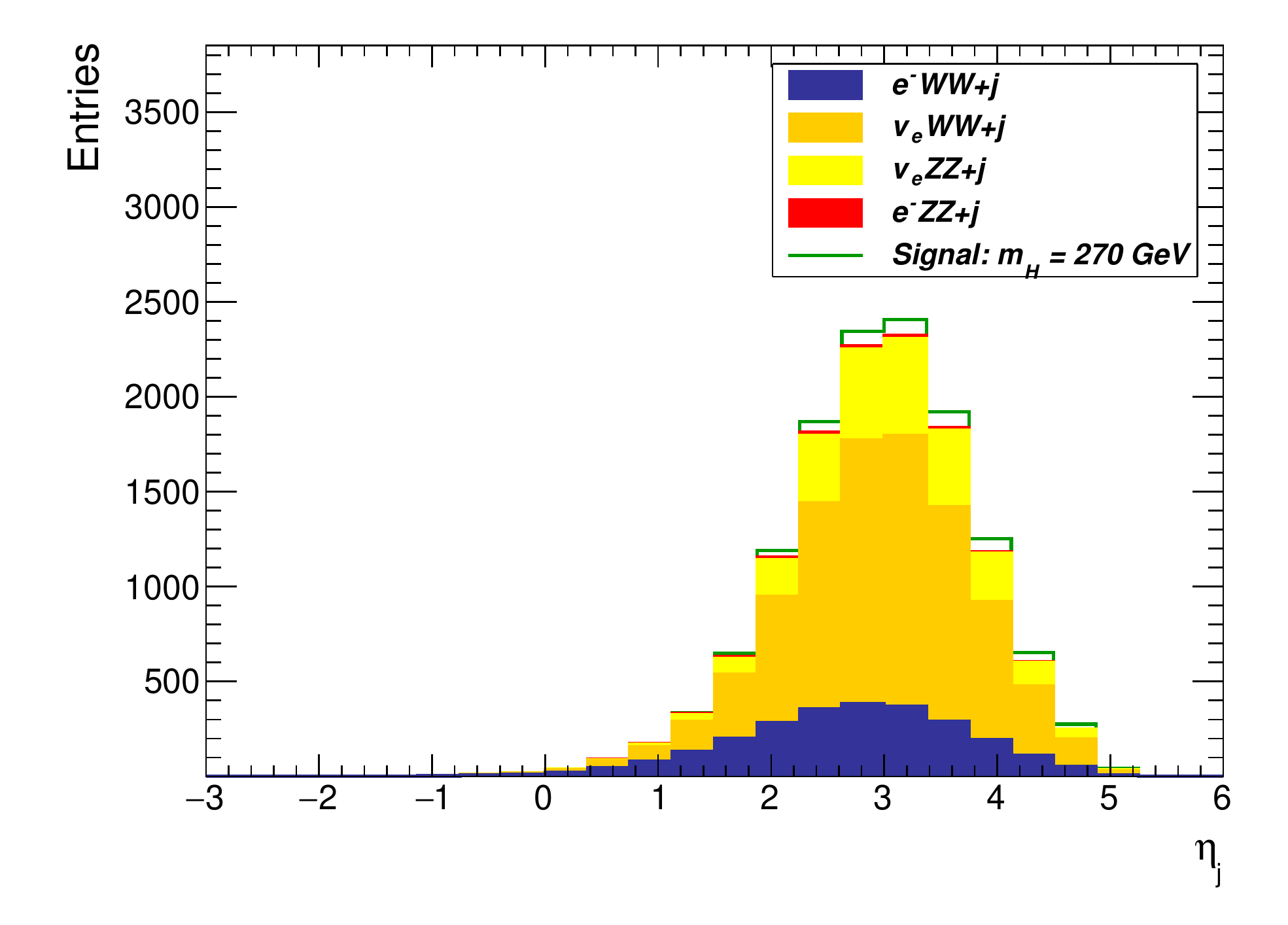}} 
\caption{(a) Multiplicity of jets in signal and backgrounds. (b) The pseudo-rapidity distribution of the forward jet after five jet selection in signal and backgrounds.}
\label{njet_eta}
\end{figure*}

\section{Event Simulation and Tools}
\label{Methodology}
The simulation of \texttt{CC} process (signal) for the heavy scalar $H$ production follows through $p e^{-} \to \nu_{e} H j$, where $\nu_e$ is electron-neutrino (and is the source of missing energy) and $j$ represents jets emanating from proton-line (we refer to this $j$ as scattered or forward jet in the text). Further the decay of $H \to W^+ W^-$ and $W^\pm \to j j$ is taken at the matrix element level for this signal process (see Fig.~\ref{fig:Feynman}). Note that $H$ can also be produced in neutral current process through the fusion of $Z$-bosons at tree-level as $p e^{-} \to e^- H j$, but the cross-section is sub-dominant {and approximately 5.5 times smaller} than the \texttt{CC} process which follows through $W^\pm$-fusion for unpolarized $e^-$ beam. 

To generate event samples for signal and potential backgrounds we use a Monte Carlo generator \texttt{MadGraph5}~\cite{Alwall:2011uj}, interfaced with a customised \texttt{Pythia-PGS}~\cite{Sjostrand:2006za} for parton showers and hadronization (for details see Ref.~\cite{Kumar:2015kca}). The detector simulation is performed using \texttt{Delphes}~\cite{deFavereau:2013fsa} with parameters optimised for the detector in LHeC. The jets are clustered using \texttt{FastJet}~\cite{Cacciari:2011ma} with the anti-$k_{T}$ algorithm~\cite{Cacciari_2008} and distance parameter $R$ = 0.4. The factorisation and renormalisation scales for the signal simulation are fixed to the heavy Higgs boson mass $m_{H}$. The background simulations are done with the default \texttt{MadGraph5} dynamic scales. The polarization of the charged electron is assumed to be $-$80\%. {This enhances the polarized cross-sections by $\sim 1.8$ times with respect to the unpolarized $e^-$ beam for both signal and background.} 

\begin{figure}[t]
\centering
    \subfloat[\label{met20}]{\includegraphics[width=0.45\textwidth]{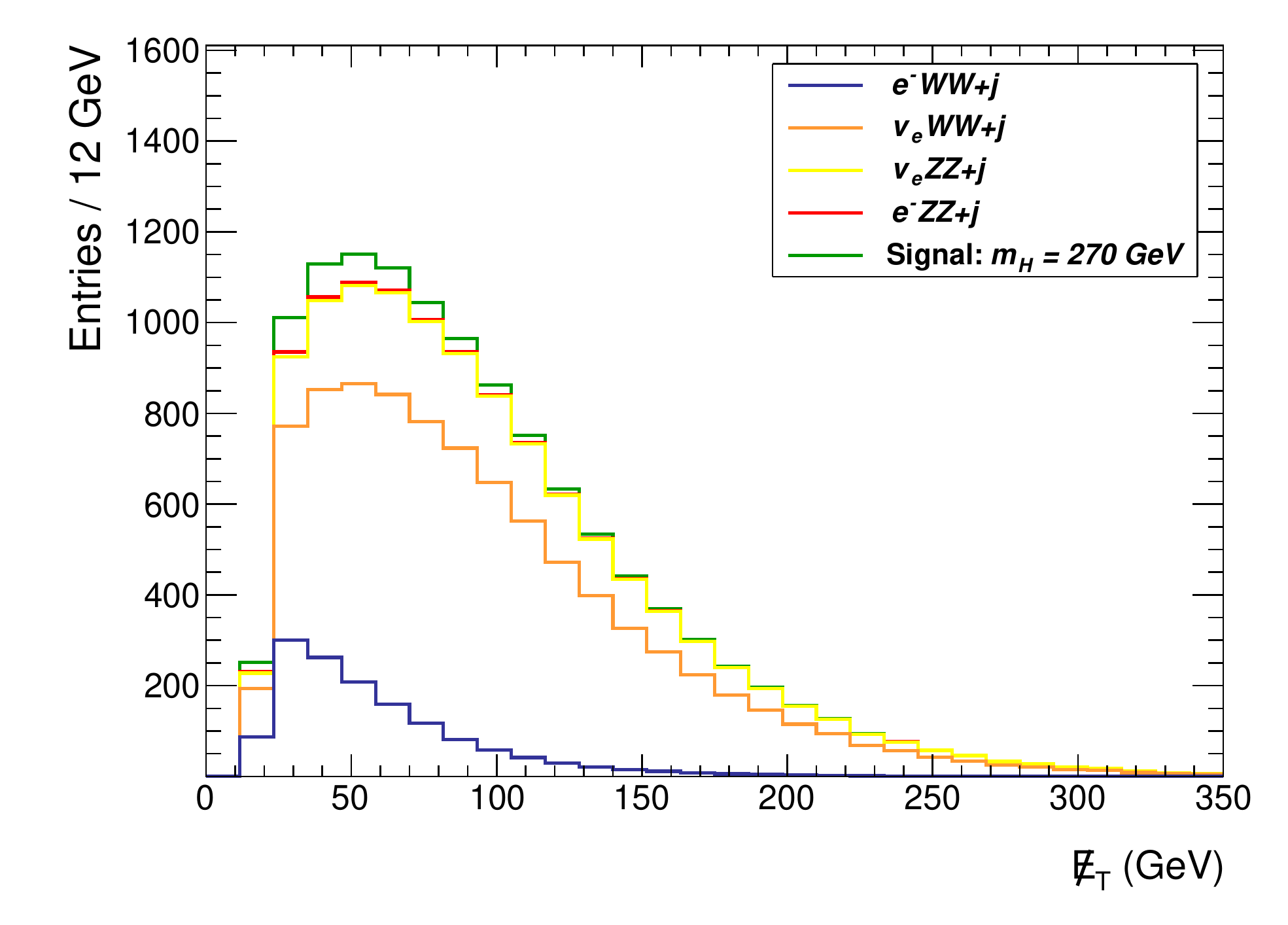}}
   \caption{The missing transverse energy distribution 
   after applying the $> 20$\,GeV requirement.}
   \label{Metplot}
\end{figure}
\begin{figure*}[t]
  \centering
    \subfloat[\label{mwp}]{\includegraphics[width=0.45\textwidth]{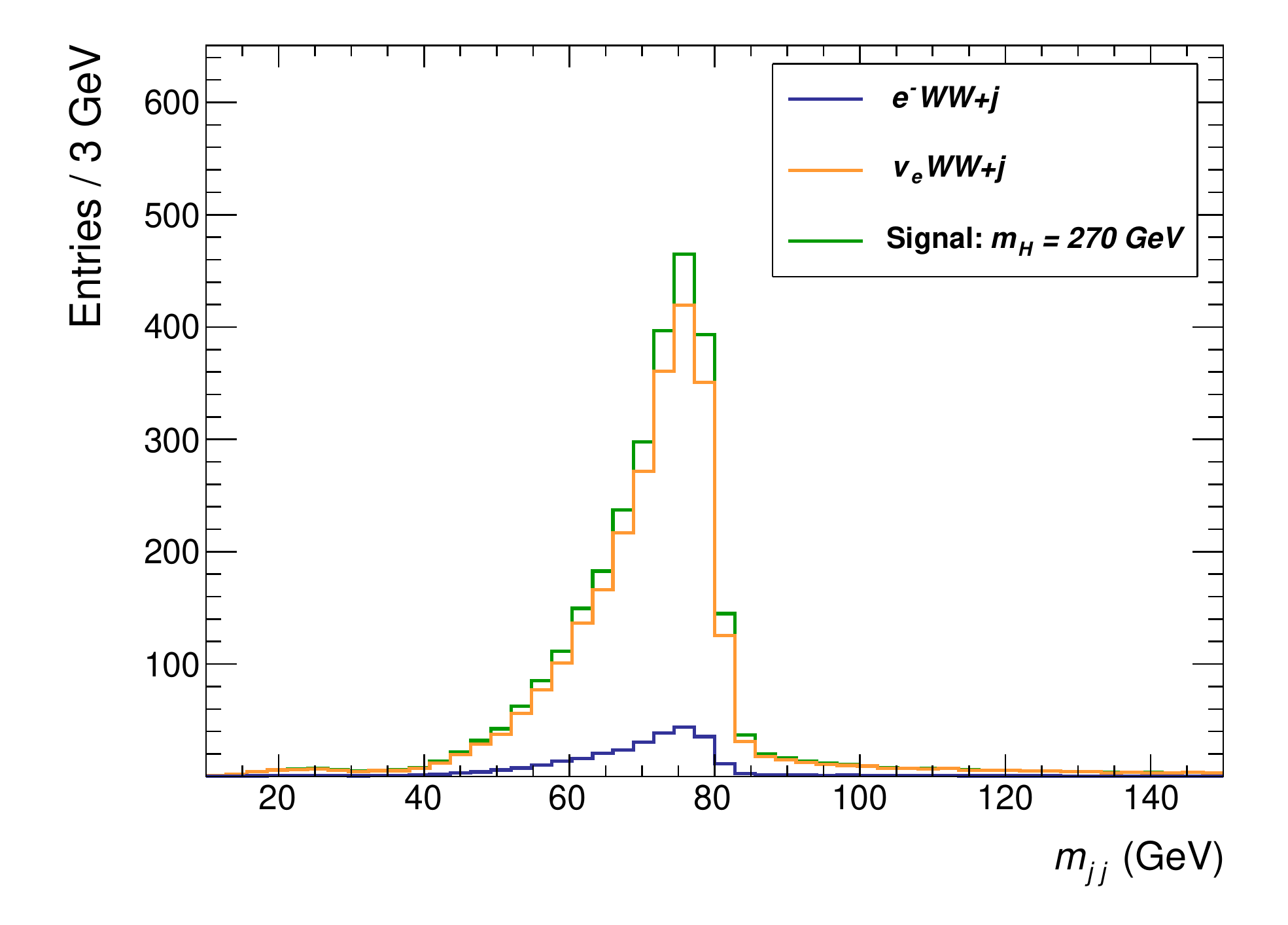}}
    \subfloat[\label{mwm}]{\includegraphics[width=0.45\textwidth]{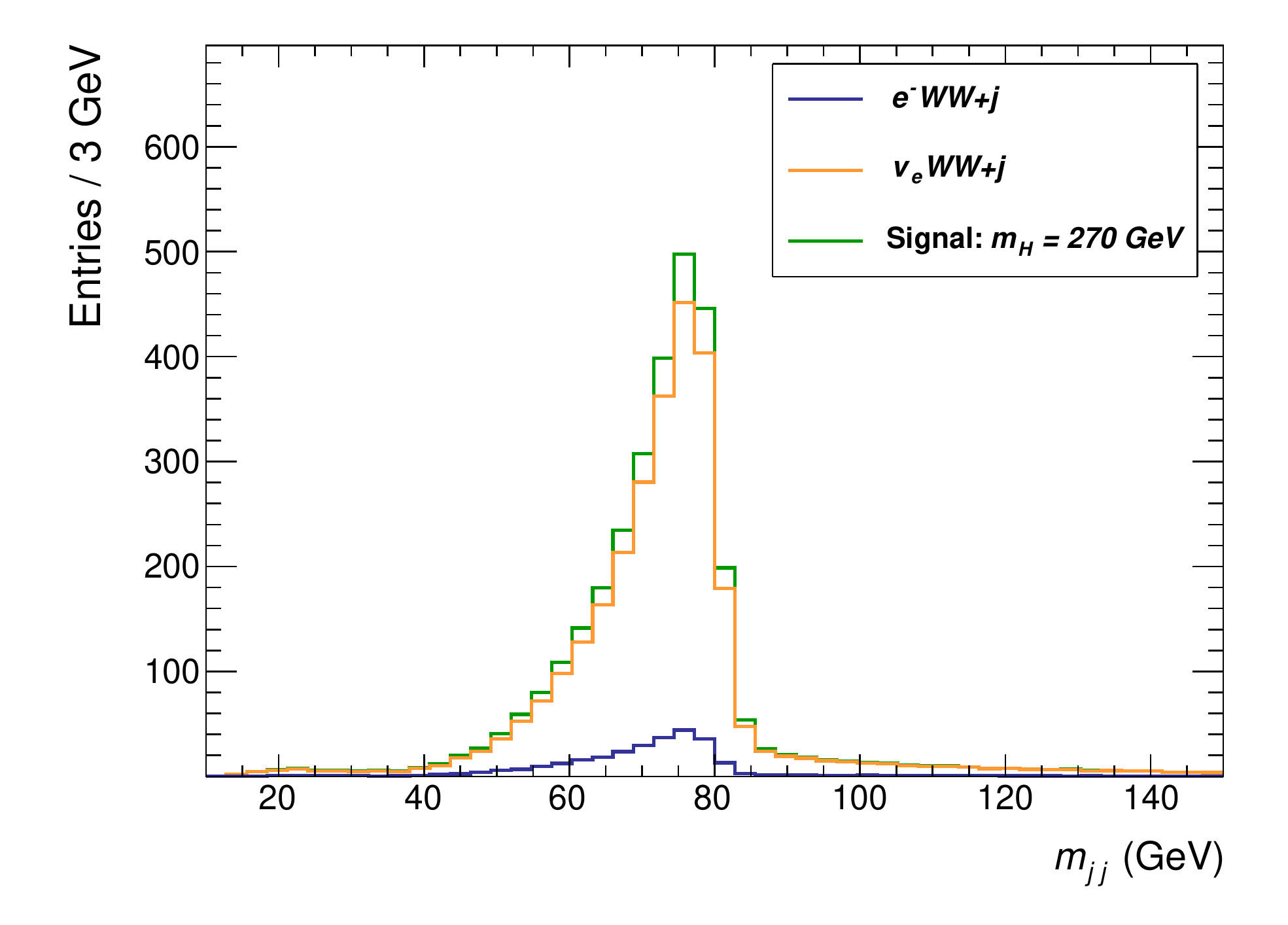}}
  \caption{Invariant di-jet mass distribution $m_{jj}$ from truth-level information of (a) $W^{+}$ and (b) $W^{-}$, where $H \to W^+ W^-$ with $m_H = 270$\,GeV.} 
\end{figure*}

An estimation of cross-section for the signal\footnote{
We scaled the $HW^+W^-$ coupling such that the cross-section for signal should be $\sim 20$ times less than the corresponding cross-section of $h$ with $m_h = 270$\,GeV. This factor is very optimistic in order to not evade any theoretical and experimental limits for $m_H$ cross-section in the considered signal.} and potential background processes are calculated at leading order using \texttt{MadGraph5} with applied minimal cuts on transverse momentum of jets $p_{T_j} > 20$\,GeV, jet pseudo-rapidity $-1 < \eta_j < 5$ and there is no requirements for transverse missing energy $E^{miss}_T$, and presented in Table~\ref{cross_section_table} for a benchmark value of $m_H = 270$\,GeV. 
Before going for mass reconstruction of $H$ with appropriate methodologies we made preliminary selection criteria to estimate the significance, and those are as follows: (a) since the final state of signal (Fig.~\ref{fig:Feynman}) contains five jets at matrix element level (four from decay of $W^\pm$-boson and one scattered jet), we chose at least five leading $p_T$-ordered jets in simulated events and (b) $E^{miss}_T > 20$~GeV. In Table~\ref{opt} we presented the number of weighted events of signal ($S$) and backgrounds ($B$) at luminosity {$\cal{L}$} = 1\,ab$^{-1}$ after these selection criteria where in the last column significance of signal over background is calculated with formula $\sigma = S/\sqrt{B}$. 
It is interesting to note the there is slight increase ($\approx 2.3$\%) in $\sigma$ after the selection of five leading jets, though $E^{miss}_T > 20$\,GeV reduces the $\sigma$ by $\approx 7$\% in comparison with initial weighted events. 
{ In order to estimate the systematic errors in the shape of signal and background distributions due to detector resolution, $E_T^{miss}$ measurement, reconstruction efficiency etc., as well as on the expected number of events we calculate significance as function of systematic factor $\delta_{sys}$: $\sigma (\delta_{sys}) = S/\sqrt{B+(\delta_{sys}\cdot B)^2}$ and added the estimation in Table~\ref{opt}.}

It is important to investigate and account for these observations during the mass reconstruction procedure of $H$ and further discuss in next section.   

\begin{figure*}[t]
  \centering
    \subfloat[\label{mljets}]{\includegraphics[width=0.45\textwidth]{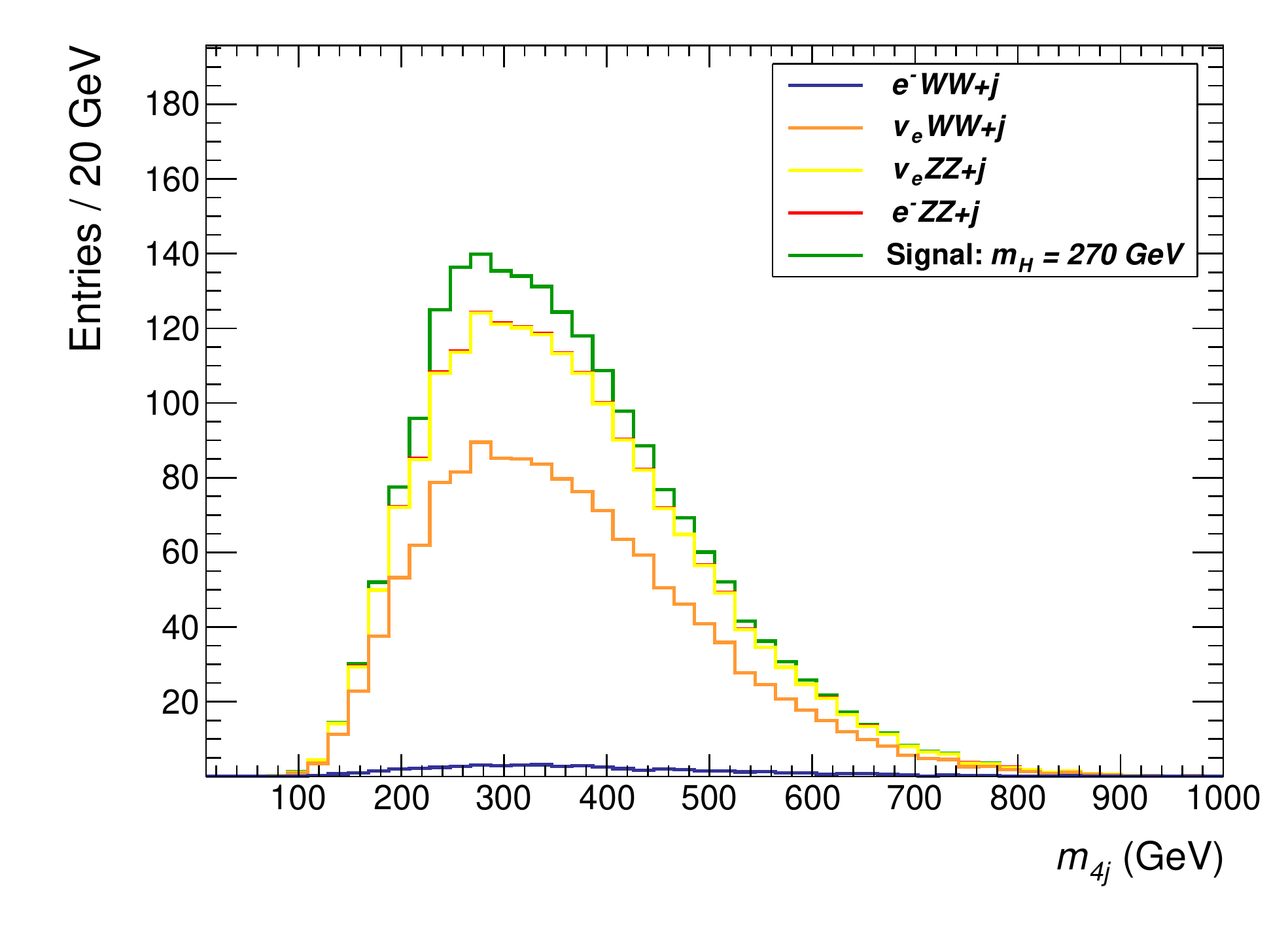}} 
    \subfloat[\label{mleta}]{\includegraphics[width=0.45\textwidth]{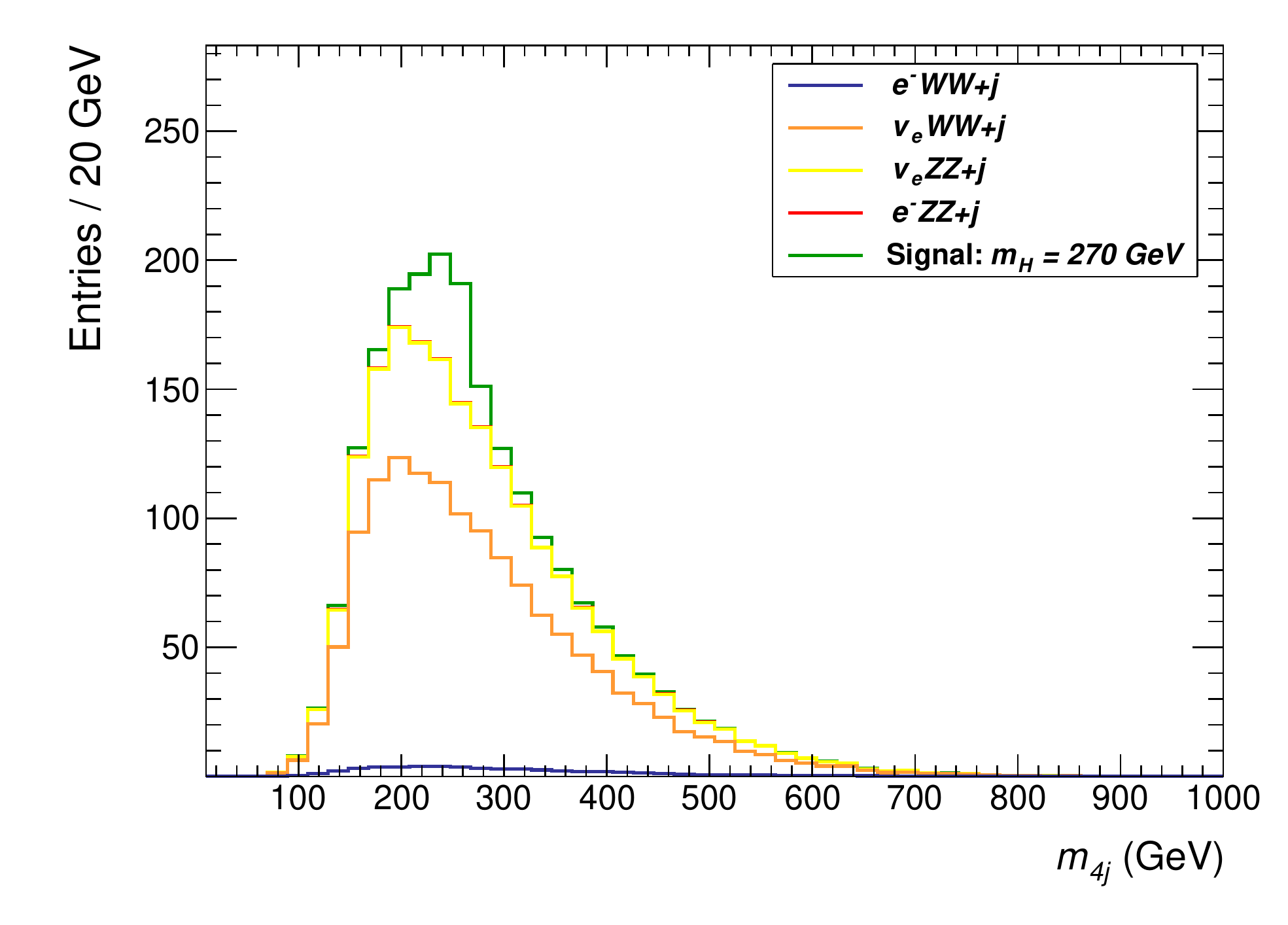}} 
  \caption{ (a) Invariant mass distribution of four $p_T$-ordered leading jets (Method\,1 (\ref{fourldjets})). (b) Invariant mass distribution of four $p_T$-ordered jets by removing the forward jet (Method\,2 (\ref{fwdjets})).}
\end{figure*}

\begin{table*}[t]
\centering
 \begin{tabular}{llllll}
 \toprule
  & Method\,1 & Method\,2 & \multicolumn{3}{c} {Method\,3} \\\cmidrule(lr){4-6}
  &  &  & BDTG & DNN & LD \\
 \toprule 
 initial &   &             & & & \\
 $\sigma_{m_{4j}}$  & 3.9$\sigma$ & 3.9$\sigma$ & 3.9$\sigma$ & 3.9$\sigma$ & 3.9$\sigma$\\
 $\sigma_{max} (m_{4j} \in [m_{4j}^{min}, m_{4j}^{max}])$   & 4.0$\sigma$ $\in [190,540]$ & 5.0$\sigma$ $\in [210,280]$ & 4.0$\sigma$ $\in [210,270]$ & 4.2$\sigma$ $\in [215,270]$ & 3.9$\sigma$ $\in [225,270]$ \\
 $S~(B)$   & 257 (4145)       & 187 (1380)      & 243 (3712)  & 237 (3258) & 237 (3689)\\
 $\sigma(\delta_{sys} = 2\%)$   & 2.4$\sigma$       & 4.0$\sigma$      & 2.5$\sigma$  & 2.7$\sigma$ & 2.5$\sigma$\\
 \hline
 $E^{miss}_{T}$ > 20 GeV            &             &             &  & &\\
 $\sigma_{m_{4j}}$      & 4.7$\sigma$ & 4.7$\sigma$ & 4.7$\sigma$ & 4.7$\sigma$ & 4.7$\sigma$ \\
 $\sigma_{max} (m_{4j} \in [m_{4j}^{min}, m_{4j}^{max}])$                     & 4.9$\sigma$ $\in [190,540]$ & 6.1$\sigma$ $\in [210,280]$ & 4.8$\sigma$ $\in [205,270]$ & 4.9$\sigma$ $\in [210,270]$ & 4.8$\sigma$ $\in [220,270]$ \\
 $S~(B)$   & 222 (2088)       & 161 (691)      & 211 (1941)  & 206 (1755) & 214 (1955)\\
 $\sigma(\delta_{sys} = 2\%)$   & 3.6$\sigma$       & 5.4$\sigma$      & 3.6$\sigma$  & 3.8$\sigma$ & 3.6$\sigma$ \\
 \bottomrule 
\end{tabular}
\caption{The significance is calculated at each stage of the optimised selection criteria using $\sigma = S/\sqrt{B}$ and {$\sigma (\delta_{sys} = 2\%) = S/\sqrt{B+(\delta_{sys}\cdot B)^2}$} where $S$ and $B$ are the expected signal and background yields at a luminosity of 1\,ab$^{-1}$ respectively. Here $\sigma_{m_{4j}}$ represents the significance in full available range in $m_{4j}$. And $\sigma_{max} (m_{4j})$ is the range where maximum $\sigma$ can be achieved, corresponding minimum to maximum range $m_{4j} \in [m_{4j}^{min},~m_{4j}^{max}]$ are specified for each approach (corresponding $S$ and $B$ are given in the next row).}
\label{method_sig}
\end{table*} 

\section{Reconstruction of the invariant mass}
\label{massrec} 

In order to reconstruct $m_H$ it is important to select appropriate hadronic jets in our signal and observe the features with respect to the dominant backgrounds. To begin the procedure we must isolate and identify the hadronic jets after detector simulations. In Fig.~\ref{njet}, number of hadronic jets are shown which are constructed with requirement on $\Delta R = 0.4$.\footnote{The distance parameter $\Delta R$ between any two particles is defined as: $\Delta{R} = \sqrt{(\Delta{\phi})^{2} + (\Delta{\eta})^{2}}$, where $\phi$ and $\eta$ are the azimuthal angle and rapidity, respectively, of particles into consideration.} It is clear that the number of hadronic jets from $ZZ$ backgrounds are competitive in comparison to the signal. Also a similar feature can be observed in the pseudo-rapidity of forward jets, $\eta_j$, as shown in Fig.~\ref{fwjet}. And therefore, the $ZZ$ backgrounds needs to be optimize with the help of missing transverse energy cut $E^{miss}_T > 20$\,GeV (see Fig.~\ref{Metplot}) and corresponding significant reduction in weighted events can be seen in Table~\ref{opt}. 

{To compare the reconstructed invariant $m_H$ with the truth-level mass, the hadronic jets originated from $W^+$ and $W^-$ bosons are selected using the truth-level information (note that $W^\pm$ are decaying from $H$ in signal). An illustration of invariant mass of two-jets, $m_{jj}$, from $W^+$ ($W^-$) is shown in Fig.~\ref{mwp} (Fig.~\ref{mwm}). Note that along with signal we only showed backgrounds with $W^\pm$ final states as there is no information stored for $Z$-bosons in truth-level.} 

After analysing these observable, we apply three different methodologies to reconstruct $m_H$ in the mentioned channel and compare the significance. In Method\,1, selection of four $p_T$-ordered leading jets are considered. Method\,2 is to select four hadronic jets excluding the most forward jet (which corresponds to largest $\eta_j$), while a high-level machine learning (ML) techniques used in Method\,3.    
\subsection{Method\,1: selection of four $p_T$-ordered leading jets}
\label{fourldjets}

In this method, all jets are sorted according to the corresponding $p_T$ and the four out-of five leading ($p_T$-ordered) jets are selected from the weighted signal and background events. We expect an inherent uncertainty in this method from the forward jet (which may not originate from either of $W^+$ or $W^-$) and this may contaminate the reconstruction of $m_H$ in the signal. The invariant mass distribution of four selected jets, $m_{4j} \equiv m_H$, using this method is shown in Fig.~\ref{mljets}. The corresponding significance $\sigma$ are shown in Table~\ref{method_sig} (second column). Here $\sigma_{m_{4j}}$ represents the significance in full available range in $m_{4j}$, and $\sigma_{max}$ is the range where maximum $\sigma$ can be achieved. This method results maximum of $4.0 \sigma$ within the invariant mass-range of $m_{4j}\in [190, 540]$\,GeV and the improvement from full range of $m_{4j}$ is by 2.5\% with initial events. However after selecting $E^{miss}_T > 20$\,GeV, accuracy of measurement improves with $4.9 \sigma$ in $m_{4j}\in [190, 540]$\,GeV (4.1\% improvement from full range). And an improvement of $\sim 16$\% in comparison with significance shown in Table~\ref{opt}.

From distribution of $m_{4j}$ in Fig.~\ref{mljets} it is noticed that the width of invariant mass is wide and reason for this could be the contamination of forward jets as discussed. Thus a method to narrower the width suppose to result better mass reconstruction by removing the forward jet and discussed in next subsection.

\subsection{Method\,2: elimination of forward jet}
\label{fwdjets}

As Method\,1 slightly improved the accuracy in the measurement of $m_H$ through four $p_T$-ordered leading jets using $m_{4j}$ (comparing the significance obtained in Table~\ref{opt}), we employ a second approach where forward jet corresponding to largest $\eta_j$ are eliminated and remaining four $p_T$-ordered jets are selected. In addition we also verified that the selected jets originate from $W^\pm$-bosons using the truth-level information. The corresponding invariant mass distribution is shown in Fig.~\ref{mleta}. Clearly the $m_{4j}$ distribution has narrower width comparing with Method\,1 (Fig.~\ref{mljets}) and this approach should improve the accuracy of measuring $m_H$. This approach also uses the same number of initial weighted events as the above method. 
When reconstructing the invariant mass of $H$, this method achieved a maximum significance of 5.0$\sigma$ before applying the missing energy cut. A maximum significance of $6.1\sigma$ can be attained with 24\% improvement after selecting jets with $E^{miss}_{T}$ > 20\,GeV. In Table~\ref{method_sig} (third column) significance obtained for Method\,2 is shown. Overall applying this method shows improvement in significance of about 33\% in comparison with significance obtained selecting at least $5j$ with $E^{miss}_T > 20$\,GeV as in Table~\ref{opt}. 
\begin{figure*}[t]
\centering
    \subfloat[\label{bdtg}]{\includegraphics[width=0.45\textwidth]{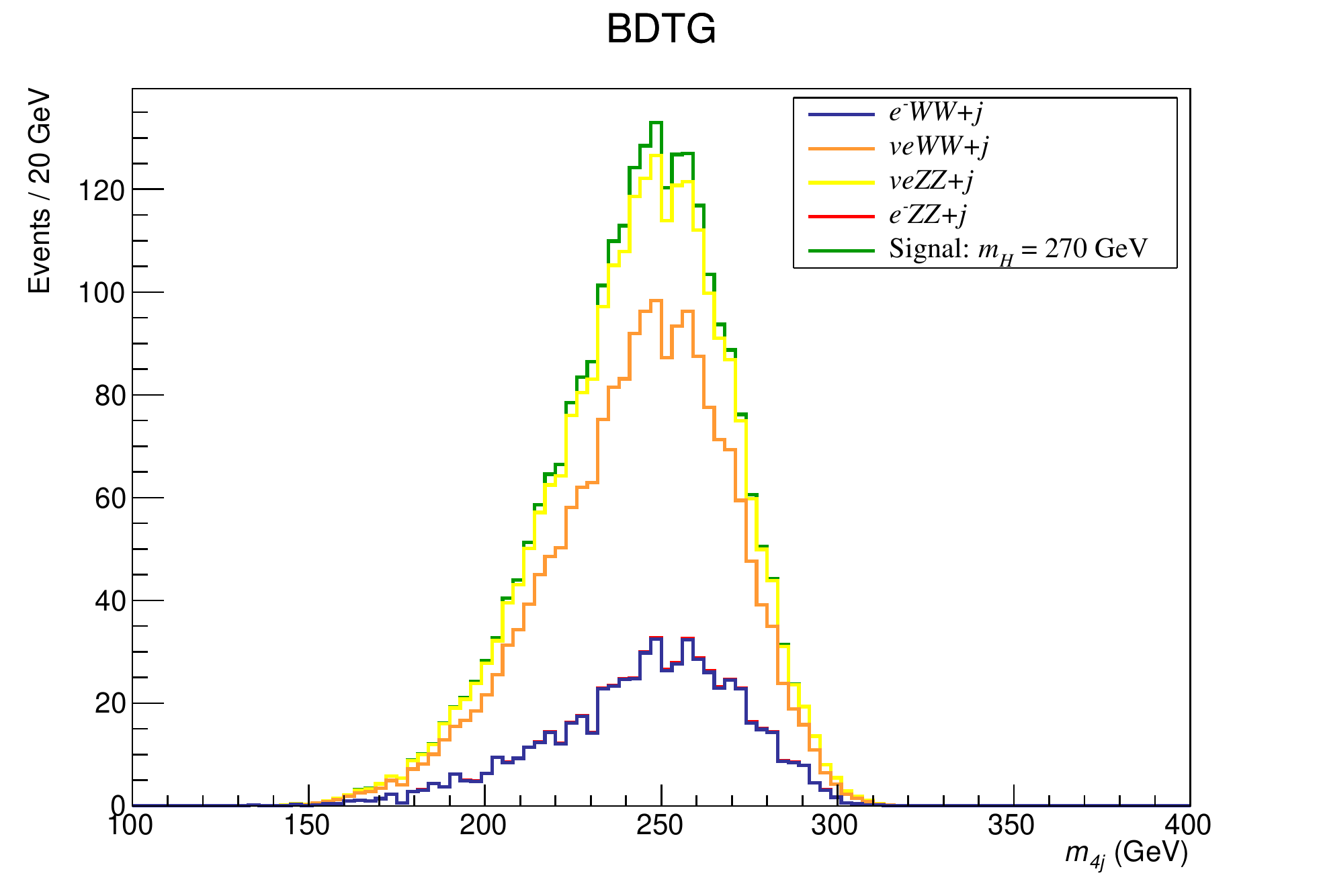}}
    \subfloat[\label{dnn}]{\includegraphics[width=0.45\textwidth]{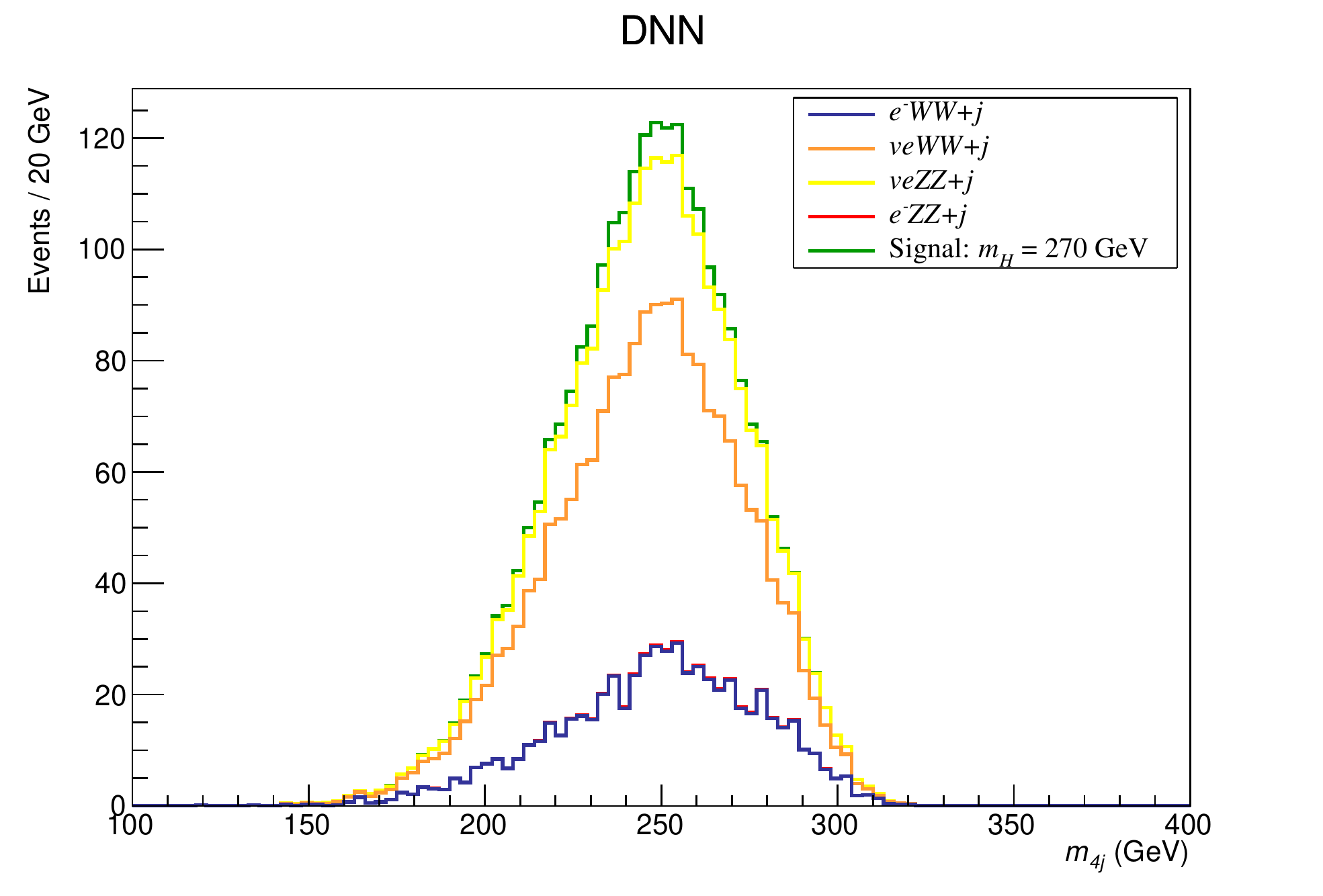}}\\
    \subfloat[\label{ld}]{\includegraphics[width=0.45\textwidth]{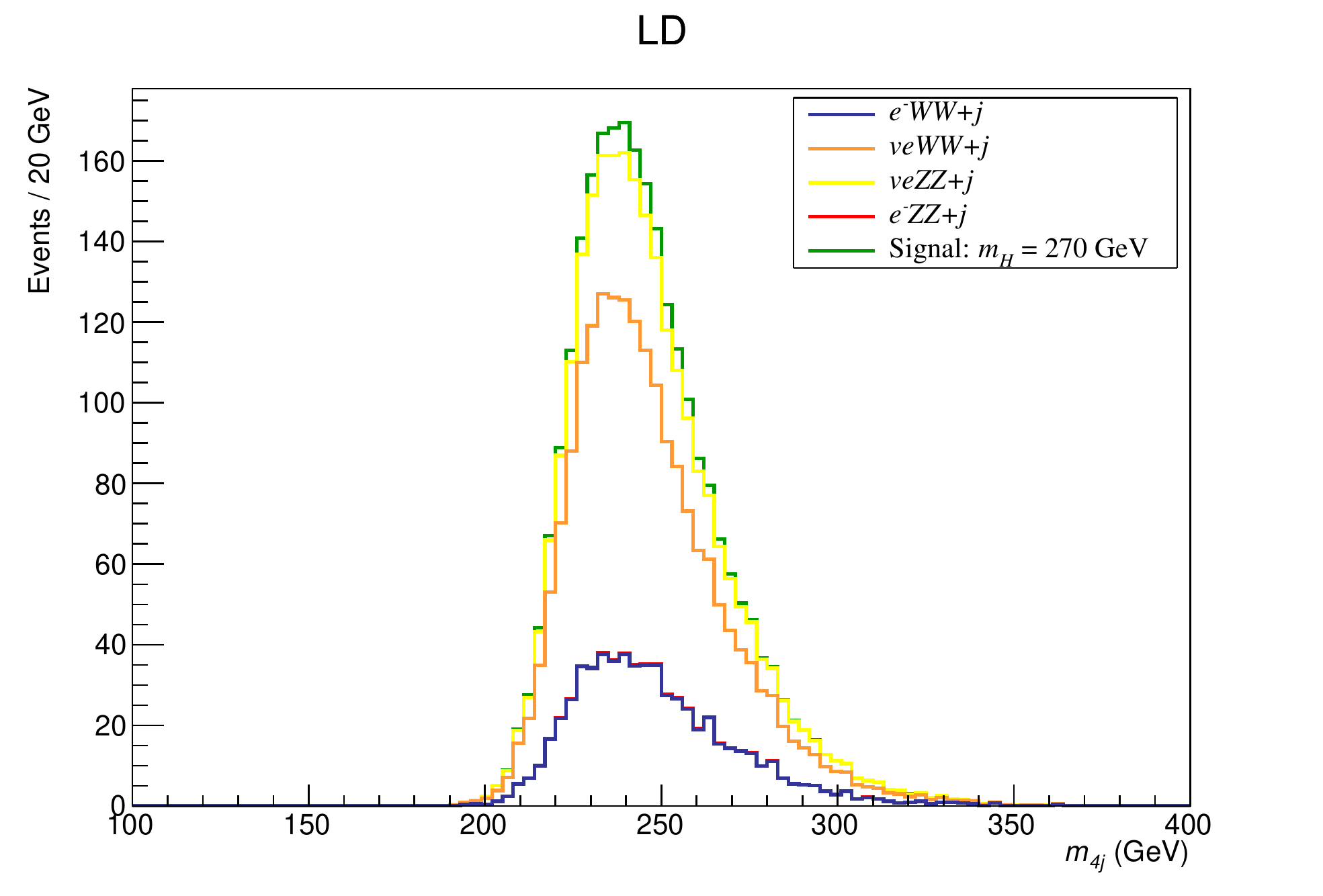}}
  \caption{Invariant mass distribution $m_{4j}$ of the trained signal and evaluated background sample using the BDT, DNN and LD method.}
  \label{mml}
\end{figure*}
\begin{figure*}[t]
   \centering
    \subfloat[\label{compldjets}]{\includegraphics[width=0.45\textwidth]{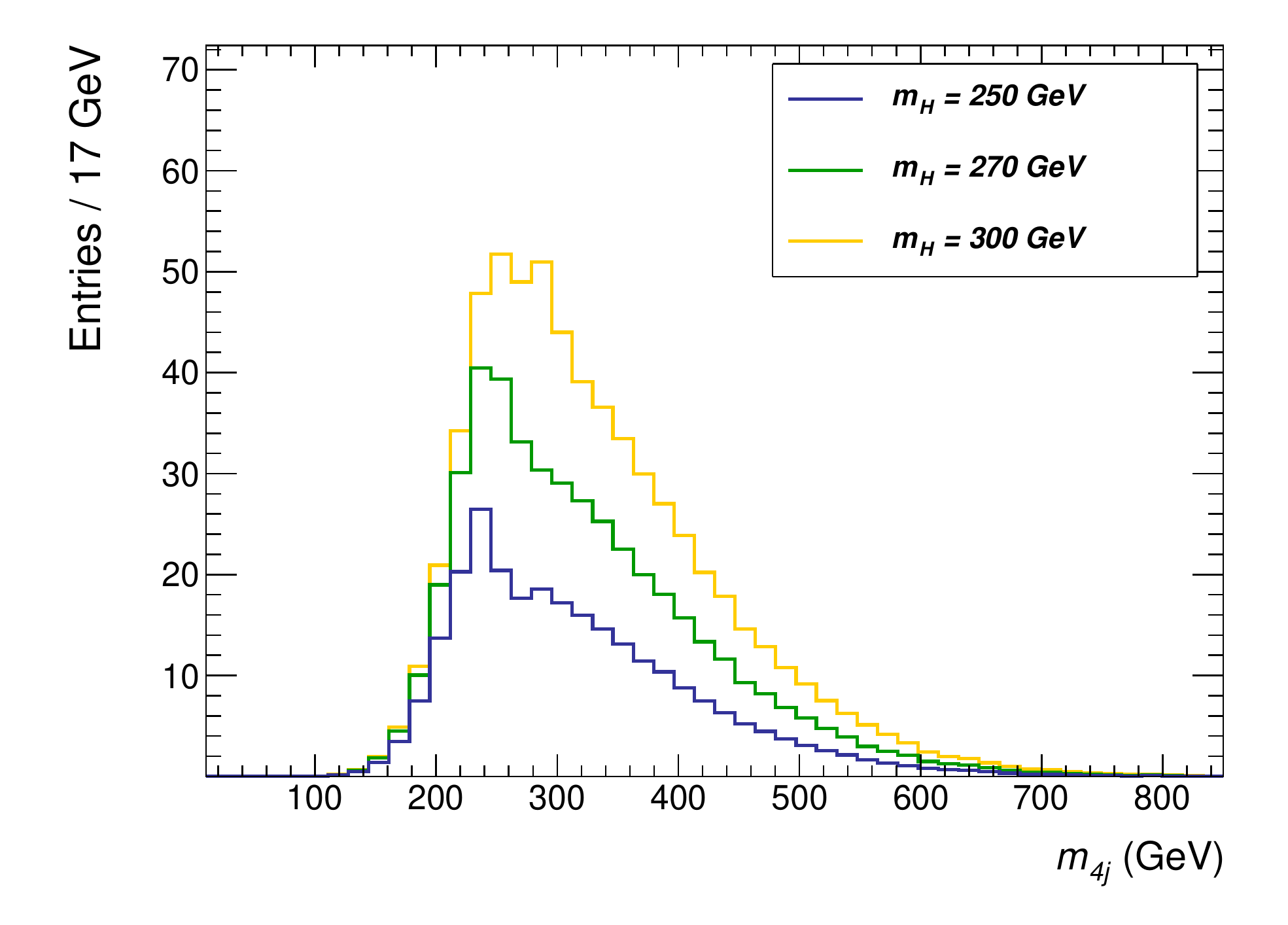}}
    \subfloat[\label{competajets}]{\includegraphics[width=0.45\textwidth]{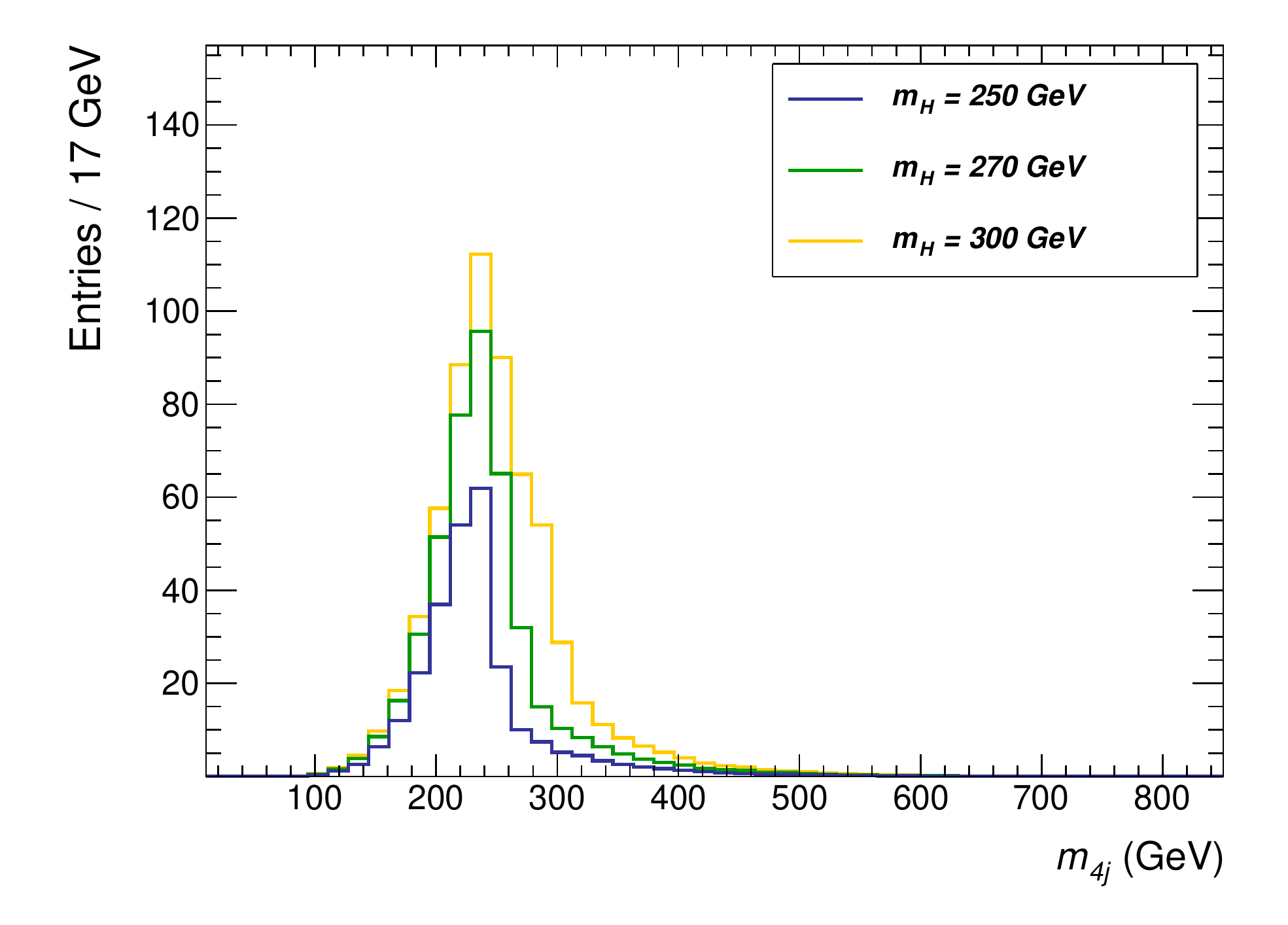}}\\
    \subfloat[\label{comDNNjets}]{\includegraphics[width=0.45\textwidth]{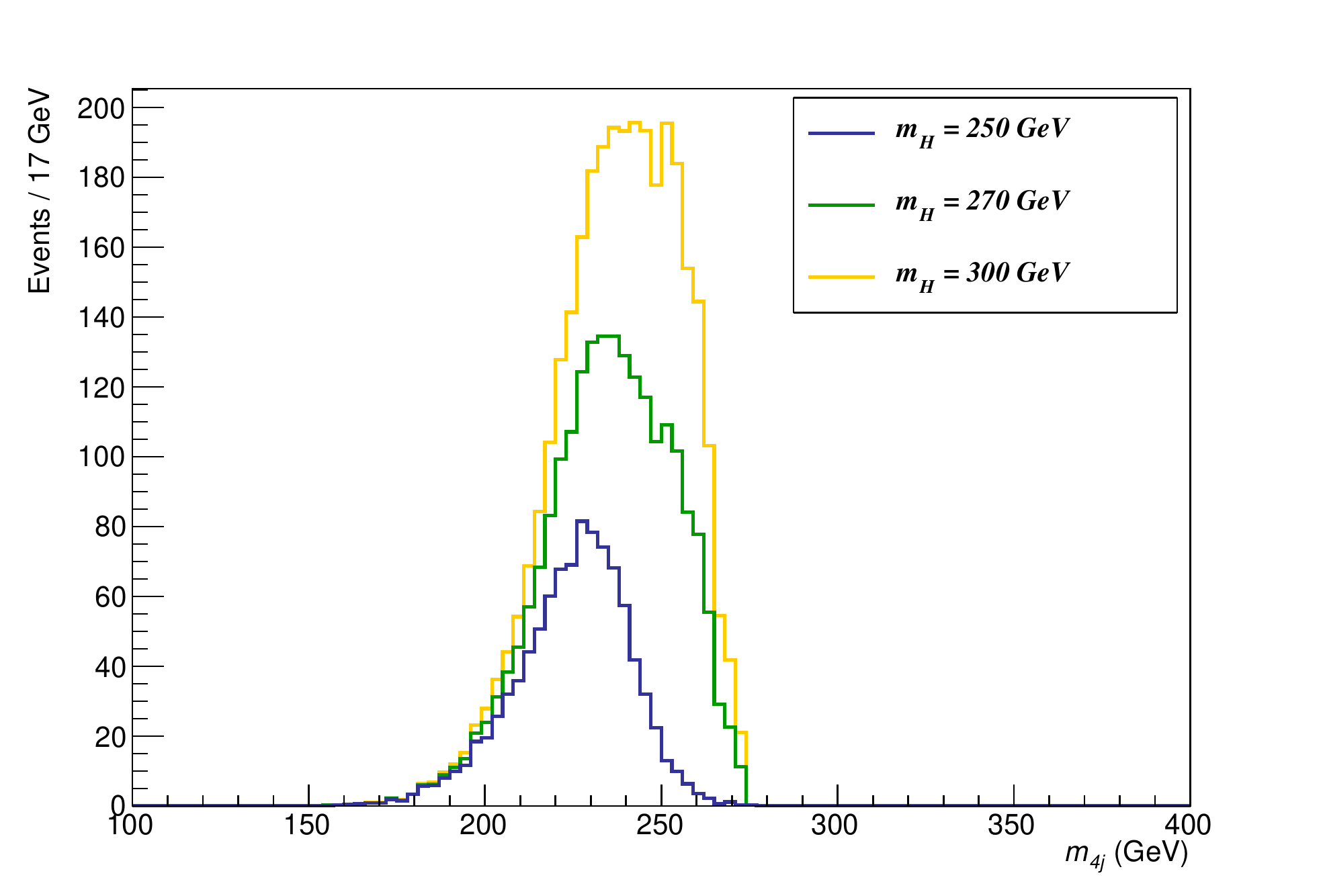}}
  \caption{Comparison of $m_{4j}$ (signal only) for three different masses: $m_{H} = 250, 270$ and 300\,GeV following (a) Method\,1 (\ref{fourldjets}), (b) Method\,2 (\ref{fwdjets}) and (c) DNN method (\ref{MLapproach}).} 
  \label{scanHmass}
\end{figure*}

\subsection{Method\,3: machine learning technique}
\label{MLapproach} 

Though the use of Method\,2 results a higher significance of about $6\sigma$ shows the efficacy of this approach to reconstruct $m_H$, we also analyse the event samples using high-level machine learning technique as Method\,3 and compare the significance. For our analysis we employed the Toolkit for Multivariate Data Analysis (TMVA) package~\cite{Hocker:2007ht} in which all multivariate methods respond to supervised learning only, i.e., the input information is mapped in feature space to the desired outputs.

To start with, the four-momentum information of jets from the signal and backgrounds' event samples are used to construct the low-level observables like jet's transverse momenta $p_{T_j}$, pseudo-rapidity $\eta_j$, azimuthal angle $\phi_j$, energy $E_j$ and mass $m_j$. The signal samples with these observables are passed in two equal proportions for training and testing, respectively, to reconstruct $m_{4j}$. Here we include three different analysis routines known as: Boosted Decision Trees with gradient boosting (BDTG), Deep Neural Network (DNN) and Linear Discriminator (LD). The details of all three analysis procedure and mechanism are documented in Ref.~\cite{Hocker:2007ht}. All background samples are passed through evaluation with default parameters in {TMVA regression application} with Boosted Decision Trees (BDTG), Deep Neural Networks (DNN) and Linear Discriminants (LD). The combination of outputs are shown in Fig.~\ref{mml}. The default parameters are later tested and tuned to give maximum significance with target mass as $270$\,GeV.\footnote{This mass is set as data-loader and defined in testing and training sample dataset as truth mass. The target mass is the reconstructed $m_{4j}$, where the selected four jets originate from $W^\pm$ as in Method\,2.}            
In Table~\ref{method_sig}, the significance obtained through all three analysis techniques are presented. All three analysis routines provides the maximum significance of mass measurement $\sim 5\sigma$, which is a little less in comparison with Method\,2 while is similar to Method\,1. Though the improvements after $E^{miss}_T > 20$\,GeV requirement are high in comparison with Method\,1. However among the three analysis routines the DNN performance is better with maximum significance of $4.9\sigma$ in $m_{4j} \in [210, 270]$.

By analysing the $m_{4j}$ distributions shown in Fig.~\ref{mml} the ML algorithms used here seems to accumulate the signal as well as the backgrounds region towards the target mass. Though the significance are consistent with other two methods and even better than Method\,1 by using DNN as shown in Table~\ref{method_sig}. 
\begin{table*}[t]
\centering
 \begin{tabular}{llll}
 \toprule
  $E^{miss}_{T}$ > 20 GeV & Method\,1  & Method\,2 & DNN  \\
 \toprule 
 $m_H = 250$\,GeV & &\\
 $\sigma_{max} (m_{4j} \in [m_{4j}^{min}, m_{4j}^{max}])$                     & 5.5$\sigma$ $\in [160,470]$ & 7.0$\sigma$ $\in [190,250]$ & 5.8$\sigma$ $\in [170,240]$ \\
 $\sigma(\delta_{sys}=2\%)$               & 4.2$\sigma$      & 6.2$\sigma$  & 5.0$\sigma$ \\ \hline
 $m_H = 300$\,GeV & &\\
 $\sigma_{max} (m_{4j} \in [m_{4j}^{min}, m_{4j}^{max}])$                     & 3.9$\sigma$ $\in [220,580]$ & 5.0$\sigma$ $\in [240,310]$ & 4.1$\sigma$ $\in [237,310]$ \\
 $\sigma(\delta_{sys}=2\%)$               & 2.9$\sigma$      & 4.4$\sigma$  & 3.1$\sigma$ \\ \hline
 $m_H = 270$\,GeV & &\\
 $\sigma_{max} (m_{4j} \in [m_{4j}^{min}, m_{4j}^{max}])$                     & 4.9$\sigma$ $\in [190,540]$ & 6.1$\sigma$ $\in [210,280]$ & 4.9$\sigma$ $\in [210,270]$ \\
 $\sigma(\delta_{sys}=2\%)$               & 3.6$\sigma$      & 5.4$\sigma$  & 3.8$\sigma$ \\ 
 \bottomrule 
\end{tabular}
\caption{Same as Table~\ref{method_sig} for $m_H = 250$ and 300~GeV in comparison with $m_H = 270$~GeV.}
\label{method_sig_scan}
\end{table*}

\subsection{Scanning $m_H$}
\label{scan}
Among the three methods, the Method\,2 - elimination of forward jet corresponding to the largest $\eta_j$ is the most efficient to reconstruct the $m_H$. So we will use this technique for two different $m_H =$ 250 and 300\,GeV, and compare the significance with the benchmark $m_H = 270$\,GeV taken in this study to understand how other masses affect the sensitivity of measurement method(s). This will allow us to investigate such masses at LHeC with considered $\sqrt{s} \approx 1.3$\,TeV. For completeness we also analyse and compare the significance with Method\,1 and DNN routines (as this method gives highest significance in comparison to BDTG and LD).

In Fig.~\ref{compldjets}, Fig.~\ref{competajets} and~\ref{comDNNjets} we compare $m_{4j}$ (signal only) using Method\,1, Method\,2 and DNN routines for $m_H = 250$,\, $270$ and $300$\,GeV, respectively. In Table~\ref{method_sig_scan}, the maximum significance obtained using Method\,1, Method\,2 and DNN are shown as in Table~\ref{method_sig}. A comparison with $m_H = 270$\,GeV shows $\sim 1\sigma$ difference in significance for both masses. Since the cross-section of $m_H = 250~(300)$\,GeV is higher (lower) than the corresponding cross-section of $m_H = 270$\,GeV, the enhancement (suppression) in significance is expected.

\section{Discussion and summary}
\label{Conclusion}

The existence of heavy particles are usually known in physics BSM and strategies to search such particles in colliders are very important. Specially in the scalar-sector it is most important since these particles are responsible for mass generation of several bosons and fermions in SM as well in BSM. In this article we attempted to prescribe mass reconstruction methods for a heavy scalar boson in a mass range of $m_H \in (2\,m_h, 2\,m_t)$, where $H$ particularly decays to hadronic jets through $W^\pm$ and the production is followed through charged-current in the LHeC environment.    

As a benchmark, a heavy scalar of mass $m_H = 270$\,GeV produced in {\texttt{CC}} channel in LHeC with $E_e = 60$\,GeV and $E_p = 7$\,TeV. Further we considered $H \to W^+W^-$ and $W^\pm \to jj$ channel to develop  a prescription for mass reconstruction. In doing so we explained the possible methods of selecting final state hadronic jets as the scattered jets in this channel are the source of contamination. Overall Method\,2 gives a significance of about $6\sigma$ using $m_{4j}$, which is better compared to the other two methodologies discussed. It is also noted that $E^{miss}_T > 20$\,GeV plays a significant role to improve the significance only when a proper selection of four hadronic jets are taken out of at least five jets. Similarly, a significant results for mass reconstruction of $m_H = 250$ and 300\,GeV with $7\sigma$ and $5\sigma$, respectively, indicates the efficiency to discover such heavy masses at future LHeC. {By accounting for the systematics effect of 2\% mentioned, the significance reduces from 6.1$\sigma$ to 5.4$\sigma$ for $m_H = 270$~GeV in Method\,2.}      


{{\it Future opportunities}: A similar analysis can be performed with $H \to ZZ$, $Z \to \ell^+\ell^-, jj$ in addition with the neutral current channel $p e^- \to e^- H j$. Also, these studies can be carried forward in the HL-LHC and proposed FCC facilities.}

\acknowledgement 
The Institute for Collider Particle Physics is grateful for the support from the National Research Foundation, the National Institute of Theoretical Physics and the Department of Science and Technology through the SA-CERN consortium and other forms of support.

\bibliography{apssamp.bib} 
\bibliographystyle{ieeetr}

\end{document}